\documentclass[aps,showpacs,prl,twocolumn]{revtex4}
\usepackage{graphicx}

\begin{document}

\title{Pressure-Induced Superconducting State of Antiferromagnetic CaFe$_2$As$_2$}
\author{Hanoh Lee$^1$, Eunsung Park$^2$, Tuson Park$^{1,2}$, V. A. Sidorov $^3$, F. Ronning$^1$, E. D. Bauer$^1$, and J. D. Thompson$^1$}
\affiliation{$^1$Los Alamos National Laboratory, Los Alamos, New Mexico 87545, USA\\ $^2$ Department of Physics, Sungkyunkwan University, Suwon 440-746, Korea\\ $^3$ Vereshchagin Institute for High Pressure Physics, Troitsk, Russia}
\date{\today}

\begin{abstract}
The antiferromagnet CaFe$_2$As$_2$ does not become superconducting when subject to ideal hydrostatic pressure conditions, where crystallographic and magnetic states also are well defined. By measuring electrical resistivity and magnetic susceptibility under quasi-hydrostatic pressure, however, we find that a substantial volume fraction of the sample is superconducting in a narrow pressure range where collapsed tetragonal and orthorhombic structures coexist. At higher pressures, the collapsed tetragonal structure is stabilized, with the boundary between this structure and the phase of coexisting structures strongly dependent on pressure history. Fluctuations in magnetic degrees of freedom in the phase of coexisting structures appear to be important for superconductivity.
\end{abstract}
\pacs{74.20.Mn, 74.25.Fy, 74.25.Dw, 74.62.Fj}
\maketitle

Discovery of superconductivity in the FeAs-layered compounds RO$_{1-x}$F$_x$FeAs (R=La, Nd, Pr, Gd, Sm) has attracted interest because of their high superconducting transition temperatures, which appear to be well outside expectations of a conventional electron-phonon pairing mechanism~\cite{kamihara08, chen08, ren08a, ren08b, liu08, mazin08}. Soon after reports of superconductivity in this 'R1111' family of compounds~\cite{kamihara08}, another family of the Fe-As superconductors, AFe$_2$As$_2$ (A=Ca, Sr, Ba, Eu), was discovered in which a non-magnetic to antiferromagnetically ordered and tetragonal (T) to low-temperature orthorhombic (O) structural transition take place simultaneously above 100~K~\cite{rotter08, ni08, yan08, krellner08, chen08b, ronning08, wu08, ni08b}. Superconductivity in the A122 family can be induced by electronic doping on the A and Fe sites~\cite{rotter08b,sefat08} or by applied pressure~\cite{park08, milton08, sebastian08}. Because of the relatively low pressures involved, the structural, magnetic and superconducting transition temperatures in CaFe$_2$As$_2$ have been explored most extensively as a function of pressure.\cite{park08, milton08,goldman08, kreyssig08, goko08, YuarXiv} These experiments led initially to conflicting conclusions about the  temperature-pressure phase diagram of CaFe$_2$As$_2$. This controversy was resolved by realizing that slight non-hydrostatic conditions lead to superconductivity; whereas, measurements in a more hydrostatic environment provided by a liquid helium pressure medium find no evidence for bulk superconductivity.\cite{YuarXiv} Under these conditions, phase boundaries delineating transitions from high-temperature tetragonal to low-temperature orthorhombic or collapsed tetragonal structures remain sharp and well-resolved as a function of pressure, in contrast to experiments with quasi-hydrostatic pressures. The higher pressures necessary to induce superconductivity in A=Sr and Ba prevent their study under ideal, hydrostatic conditions, with the result that their resistively determined pressure-temperature phase diagrams show substantial variability depending quasi-hydrostatic conditions.~\cite{sebastian08,kumar08a,kotegawa09,igawa09,fukazawa08,mani09} Whether these compounds are similar to CaFe$_s$As$_2$ in not supporting superconductivity under hydrostatic pressure remains unknown.

Though the observation of no superconductivity in CaFe$_2$As$_2$ under liquid helium pressures is important for an interpretation of the physics of this material, the appearance of superconductivity with quasi-hydrostatic pressures also raises the interesting question of what is so different in these two cases. From  measurements of the electrical resistivity, magnetic susceptibility and strain as a function of pressure generated in a liquid pressure medium, we find that the temperature-pressure phase diagram of CaFe$_2$As$_2$ depends strongly on pressure history and that the superconducting volume fraction peaks in a narrow pressure range. As discussed below, these observations suggest that superconductivity emerges as a result of fluctuations associated with the complex state that appears in the pressure interval between approximately 0.3 and 0.8~GPa.

Plate-like single crystals of CaFe$_2$As$_2$, which crystallize in the ThCr$_2$Si$_2$ tetragonal structure, were grown from a Sn flux.\cite{ronning08} Electrical resistivity measurements were performed using a conventional four-probe technique with an LR-700 resistance bridge, ac magnetic susceptibility measured at 157~Hz using a Stanford Lock-in Amplifier SR830, and strain determined from the resistance of a 12-$\mu$m diameter constantan gauge glued along the tetragonal a-axis. Resistivity and ac susceptibility of CaFe$_2$As$_2$ were measured simultaneously in a clamp-type Be-Cu pressure cell up to 1.52~GPa, where a silicone fluid was used to produce a quasi-hydrostatic pressure environment; whereas, strain measurements were performed separately but in the same pressure cell. The superconducting transition temperature of Pb was measured inductively to determine pressures at low temperature.\cite{eiling81} Independent studies show~\cite{klotz} that below 2 GPa the pressure gradient supported by silicone fluid is very low, provided it remains in the liquid state. Upon cooling, however, silicone fluid freezes into a structurally glassy state below a temperature T$_f$ $\approx$122+aP,  where a=87 K/GPa.\cite{sidorov} In the frozen state, the pressure gradient $\sigma \approx 0.07P$, so that at an average pressure of 1 GPa, the pressure gradient could be as large as 0.07 GPa.\cite{klotz} By using samples that fill only a small fraction of the cell volume, we expect a smaller pressure gradient across the sample. Indeed, the maximum transition width of the Pb superconducting transition corresponded to about 0.01 GPa at the highest average pressure, signalling the presence of a small ($\leq$~1\%) pressure inhomogeneity.

Several single crystals were studied  by electrical resistivity at ambient pressure; crystals with Sn inclusions showed a larger relative resistivity ratio (RRR = $\rho$(300~K)/$\rho$(0~K)~$\sim15$) and those without Sn inclusions had a smaller RRR ($\sim5$). When both types of samples were subject to pressure, however, characteristic features, such as the structural/antiferromagnetic transition temperature $T_S$ and superconducting transition temperature $T_c$, were similar to each other. Below, we report data on single crystals CaFe$_2$As$_2$ free of Sn inclusions. Though the ambient-pressure RRR ($\sim 5$) was small, it increased to greater than 50 at high pressure, indicating intrinsically high crystallinity with few defects.

\begin{figure}[tbp]
\centering  \includegraphics[width=7.5cm,clip]{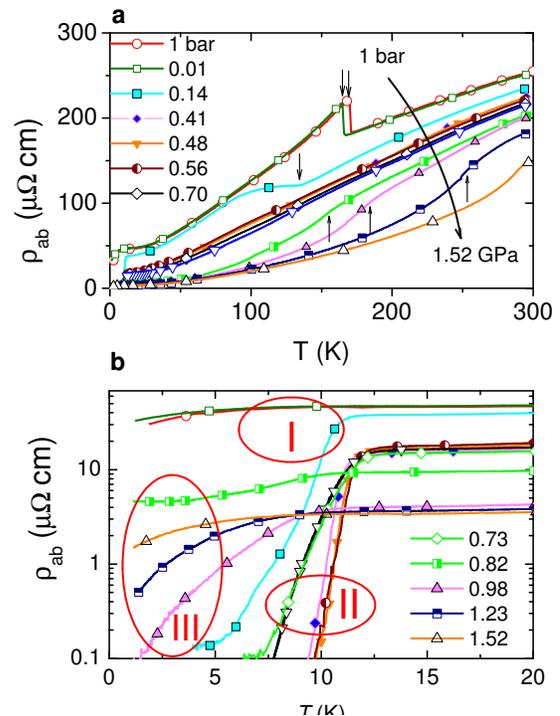}
\caption{(color online) Electrical resistivity of CaFe$_2$As$_2$ under pressure. Curves were obtained on cooling, except for P = 1~bar where measurements were made on both cooling and warming. The legends apply to both \textbf{a} and \textbf{b} and give the pressure in GPa. \textbf{a} In-plane resistivity $\rho _{ab}$ from 1.2 to 300~K for pressures of 1~bar, 0.01, 0.14, 0.41, 0.48, 0.56, 0.7, 0.73, 0.82, 0.98, 1.23, and 1.52~GPa. Arrows denote temperatures for a phase change, as discussed in the text. For clarity, only one out of 50 data points is shown. \textbf{b} An expanded view of the low-temperature, in-plane resistivity $\rho _{ab}$ shows the evolution of the superconducting transition temperature $T_c$. Sharp transitions appear only in a phase of coexisting O and cT structures. Curves corresponding to pressure regimes in Fig. 2b are circled. For clarity, one out of 10 data points is shown.}
\label{figure1}
\end{figure}
Figure~1a displays the  electrical resistivity ($\rho_{ab}$) of CaFe$_2$As$_2$ for electrical current flowing in the Fe-As plane. At ambient pressure, there is a step-like, hysteretic increase in the resistivity at 171~K (=$T_S$), where the antiferromagnetic (AFM) and lattice-structural transitions coincide.\cite{seungho} When cooled through $T_S$, the high-temperature tetragonal structure switches to an orthorhombic structure and the Fe spin is aligned along the orthorhombic a-axis with an ordered moment 0.8~$\mu_B$/Fe~\cite{goldman08}. Pressure strongly suppresses $T_S$ at an initial rate of -220~K/GPa and the transition width broadens. For pressures higher than 0.35~GPa, where neutron-diffraction measurements found a collapsed tetragonal (cT) structure at 0.63~GPa and 50~K~\cite{kreyssig08}, a signature for $T_S$ is hardly visible in the resistivity. In this higher pressure range, muon-spin resonance studies show that magnetic order persists to at least 0.62~GPa.\cite{goko08} On the basis of symmetry, magnetic order with the ambient pressure propagation wave-vector must be associated with the O-structure. With further increasing pressure ($P>0.75$~GPa), a break in the slope of the resistivity (marked by arrows in Fig.~1a) appears, and its temperature $T_S^h$ increases with increasing pressure, which signals a change to essentially phase-pure cT structure \cite{YuarXiv}. (We note that experiments using a liquid helium pressure medium find the low-temperature O-cT boundary at P $\approx 0.35~ GPa$.\cite{YuarXiv}) As shown in Fig.~2a, the transition at $T_S^h$ is strongly hysteretic with temperature, and though not apparent in these figures, the RRR jumps from $\sim 15$ to over 50 upon entering the new structural phase.

\begin{figure}[tbp]
\centering  \includegraphics[width=7.5cm,clip]{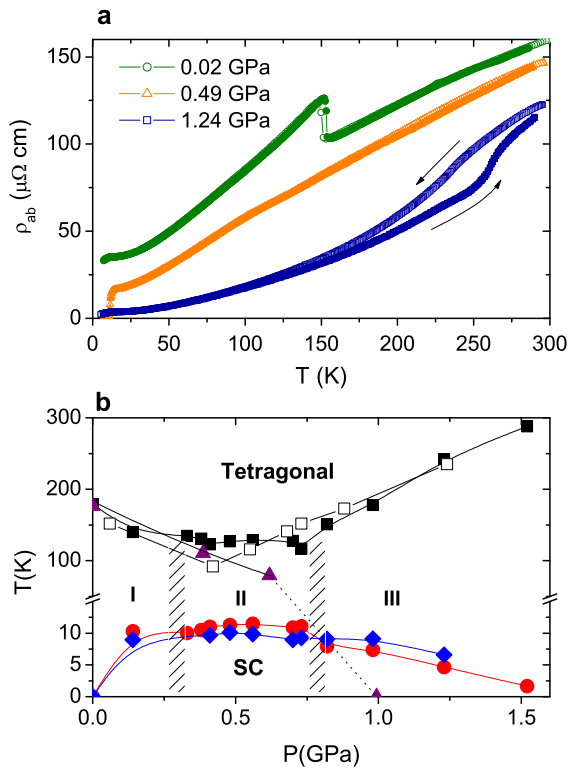}
\caption{(color online) \textbf{a} Temperature-dependent resistivity upon cooling and warming in representative pressure ranges, corresponding to those in phases I, II and III in \textbf{b}. Note the emergence of strong thermal hysteresis at $T_S^h$ in phase III. \textbf{b} Temperature-pressure phase diagram. Solid squares (increasing pressure) denote the structural transition temperature $T_S$ from a tetragonal to a orthorhombic for $P<0.3$~GPa (phase I), a phase of coexisting O and cT structures for $0.3<P<0.8$~GPa (phase II), and the cT structure (phase III) for $P>0.8$~GPa. Open squares represent structural transitions determined with decreasing pressure. The transition temperatures were defined from a break or maximum in the slope of d$\rho/$dT. Hashed vertical lines separate these structural phases. Circles denote the midpoint of the resistive SC transition and diamonds the onset of superconductivity in ac magnetic susceptibility $\chi_{ac}$. Triangles describe the magnetic transition from a paramagnetic to a SDW phase deduced from $\mu$SR measurements.\cite{goko08}}
\label{figure2}
\end{figure}
A slight downturn in the resistivity $\rho_{ab}$ of CaFe$_2$As$_2$ is observed below 8~K at ambient pressure (Fig.~1b), possibly due to disconnected superconducting filaments. At a pressure of 0.14~GPa, the resistance drops by a factor of 300 at 5.4~K from its value at the SC onset temperature (=11.5~K). With further applied pressure, the resistive transition becomes sharp in a narrow pressure range, and the mid-point temperature $T_c$ increases, goes through a maximum near 0.5~GPa and forms a wide pressure dome of $T_c$'s (circles in Fig.~2b), consistent with an earlier report from experiments with quasi-hydrostatic pressure~\cite{milton08}.  The existence of superconductivity over a wide pressure range suggests that a specific crystal structure is not a requirement for superconductivity in CaFe$_2$As$_2$, but as will be discussed, this conclusion is not supported by other measurements.

Squares in Fig.~2b denote transitions indicated by arrows in Figs.~1a; whereas, triangles in Fig.~2b represent the AFM transition ($T_N$) obtained from $\mu$SR measurements~\cite{goko08}. Below 0.3~GPa, both $T_N$ and $T_S$ are suppressed at a similar rate under pressure. For pressures higher than 0.3~GPa, where a collapsed tetragonal structure grows in, the two transitions are decoupled: $T_S$ is almost independent of pressure, but $T_N$ drops rapidly to zero Kelvin between 0.62 and 1.0~GPa. The $T-P$ phase diagram in Fig.~2b shows the coexistence of magnetism and superconductivity.
\begin{figure}[tbp]
\centering  \includegraphics[width=7.5cm,clip]{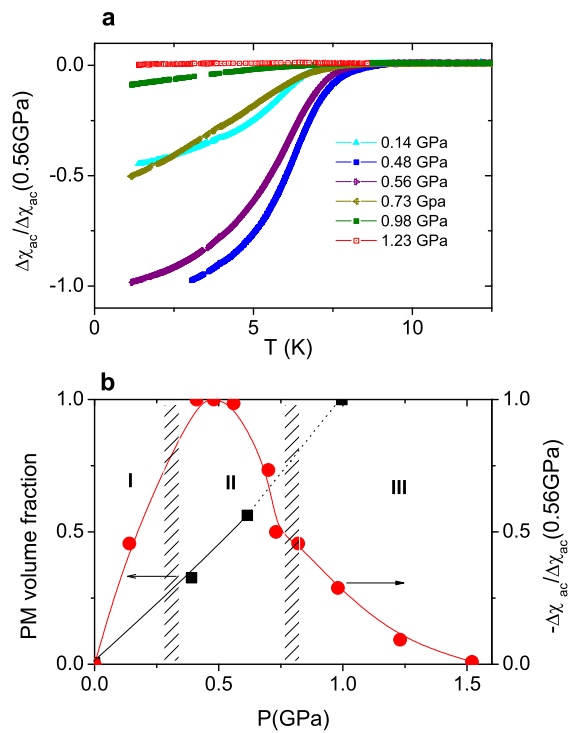}
\caption{(color online) \textbf{a} Pressure evolution of normalized ac magnetic susceptibility $\Delta \chi_{ac} / \Delta \chi_{ac}$(=0.56~GPa) on a single crystal sample, where $\Delta \chi_{ac} = \chi_{ac}-\chi_{ac}(T_c)$. In these units, 1.0 corresponds to at least $0.5(1/4\pi)$ as explained in the text. The amplitude of the ac magnetic field was approximately 1 Oe.  \textbf{b} Diamagnetic response $-\Delta \chi_{ac}(P) / \Delta \chi_{ac}$(0.56~GPa) (circles) at 1.2~K and volume fraction of a paramagnetic phase estimated by $\mu$SR measurements (squares), where the dotted line is an extrapolation~\cite{goko08}. Hashed vertical lines are the same as those shown in Fig.~2b.}
\label{figure3}
\end{figure}

The bulk nature of SC was investigated by ac magnetic susceptibility measurements, with results plotted in Fig.~3a for a single crystal sample. At a pressure close to 1~bar, there is no noticeable change in $\chi_{ac}$ (not shown) at low temperatures, which is consistent with other bulk measurements. With pressure, a drop in $\chi_{ac}$ occurs at the resistive mid-point $T_c$. The drop, reflecting the volume shielding fraction of superconductivity, initially increases with pressure, goes through a maximum between 0.4 and 0.6~GPa, and becomes negligible at 1.52~GPa (see Fig.~3b). The change in $\chi_{ac}$ of CaFe$_2$As$_2$ at 0.5~GPa is at least 50~\% of perfect diamagnetism based on measurements in the same coil with known superconductors of similar shape and mass, ruling out filamentary superconductivity for CaFe$_2$As$_2$ in this pressure range. A similar diamagnetic response has been reported in superconducting crystals of  CaFe$_{1.94}$Co$_{0.06}$As$_2$.\cite{kumar08}

Because these low-field ($\approx 1$~Oe) ac susceptibility measurements reflect diamagnetic shielding of the bulk single crystal, these measurements were repeated on crushed crystals that were annealed for 24~hr at 300~C after crushing to remove potential strain induced by grinding them into powder. Scanning electron microscopy showed that the crushed powder was composed of thin (1-2 $\mu$m-thick) platelets but with a distribution of diameters (none larger than 100~$\mu$m in diameter, with about 15\% of the particles being in each of the ranges 50-80, 30-50, 10-30, and 5-10~$\mu$m and the remaining $\approx 40\%$ having diameters of 1-5~$\mu$m). Susceptibility measurements on this powder gave the same evolution with pressure and nearly the same diamagnetic response that are shown in Fig.~3a. The superconducting penetration depth in CaFe$_2$As$_2$ is not known but is $\approx 0.45$~$\mu$m in Ba$_{.55}$K$_{.45}$Fe$_2$As$_2$ \cite{aczel} with T$_c$ = 30~K and close to this value in F-doped LaFeAsO (T$_c$ = 18~K) \cite{takeshita}. With T$_c$ in CaFe$_2$As$_2$ roughly half that in these other iron arsenide superconductors, we assume its  penetration depth is roughly $\sqrt{2}$ larger, i.e., $\approx0.6$~$\mu$m.~\cite{uemura91,aczel} Consequently, the ac magnetic field samples of order 60 to 100~\% of the thickness of the crushed crystals and a comparable volume of the large fraction of smallest crystallites. Though still not completely a probe of superconductivity in the bulk, these experiments show that a substantial volume of CaFe$_2$As$_2$ supports superconductivity.

Squares in Fig.~3b represent the pressure evolution of the paramagnetic volume fraction (the fraction that is not ordered antiferromagnetically) of CaFe$_2$As$_2$ estimated independently from $\mu$SR measurements~\cite{goko08}. Neutron-diffraction ~\cite{goldman082} and NMR measurements~\cite{seungho2} are consistent with this estimate being an upper limit on the paramagnetic volume fraction. At 0.4~GPa, the fractional diamagnetic response exceeds the paramagnetic volume fraction of the cT phase, indicating that the pressure-induced superconductivity in CaFe$_2$As$_2$ is not from a pure, phase-separated paramagnetic phase, but involves more of the sample volume.

Squares in Fig.~4a show the pressure evolution of the residual resistivity ($\rho_0$) of CaFe$_2$As$_2$, which was obtained from a least-squares fit of the low temperature resistivity to a simple power law, $\rho = \rho_0 + AT^n$. Values of $n(P)$ range between 2 and 3~(not shown), but  $\rho_0$ is relatively insensitive to the precise value of $n$. The residual resistivity displays a sharp change at 0.3 and 0.8~GPa, where O to coexisting O/cT and O/cT to cT structure changes occur, respectively. In the orthorhombic ($P<0.3$~GPa) and collapsed tetragonal phase ($P>0.8$~GPa), $\rho_0$ monotonically decreases, but the residual resistivity and $T_c$ are non-monotonic, forming a dome around 0.56~GPa in the coexisting phase $0.3<P<0.8$~GPa where the diamagnetic response is a maximum. The deviation from a monotonic decrease in $\rho_0$ signifies an additional scattering mechanism at low temperatures.

\begin{figure}[tbp]
\centering  \includegraphics[width=7.5cm,clip]{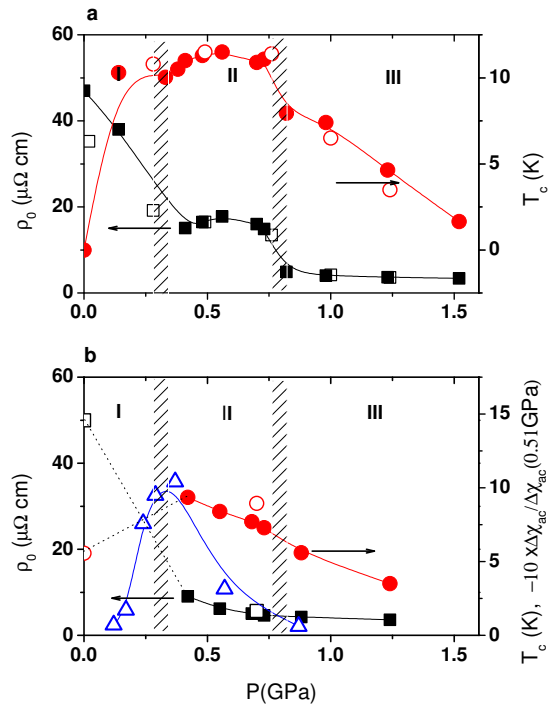}
\caption{(color online) \textbf{a} Pressure dependence of the residual resistivity $\rho_0$ (squares) of CaFe$_2$As$_2$ plotted on the left ordinate and superconducting transition temperature $T_c$ (circles) plotted on the right ordinate. Solid and open symbols are data obtained with increasing pressure, for the same crystal $(\#1)$ reported in previous figures and for a second crystal $(\#2)$, respectively. The residual resistivity of $\#2$ was normalized to that of $(\#1)$ at 1.2~GPa. Hashed vertical lines are the same as those shown in previous figures. \textbf{b} Residual resistivity and resistive mid-point $T_c$ of crystal $\#1$ (solid symbols) and for $\#2$ (open symbols) obtained on decreasing pressure from 1.52~GPa. Open triangles are the normalized diamagnetic response, defined in Fig.~3, for decreasing pressure. Hashed lines from \textbf{a} are shown for comparison.}
\label{figure4}
\end{figure}

The existence of structural and magnetic inhomogeneity in phase~II is central to the emergence of bulk superconductivity. NMR spectra obtained on CaFe$_2$As$_2$ under quasi-hydrostatic pressure clearly reflect this inhomogeneity ~\cite{seungho2} as do stain-gauge measurements shown in Fig.~5. For pressures corresponding to phase~I in Fig.~2b, the strain gauge shows a sharp change of resistance at the structural/magnetic transition temperature. In this pressure range, the pressure medium remains a liquid to just below the structural transition temperature T$_S$. The value of this resistance change at T$_S$  $\Delta R/R$ corresponds to a length change  $\Delta L/L$ = - 0.24 \%, taking into account the gauge factor of constantan $G = 2~ ( \Delta L/L \approx  G^{-1} \Delta R/R)$. This length change along the a-axis corresponds well that obtained by x-ray diffraction on CaFe$_2$As$_2$.\cite{ni} On the other hand, for measurements in phase~II, where the freezing temperature of the pressure medium exceeds T$_S$, the length change becomes distributed over a very broad temperature range below the onset of the transition, which implies a wide range of coexisting paramagnetic phase and antiferromagnetic orthorhombic phase. This also is consistent with a microscopic probe of the structure by NMR.\cite{seungho2}

\begin{figure}[tbp]
\centering  \includegraphics[width=7.5cm,clip]{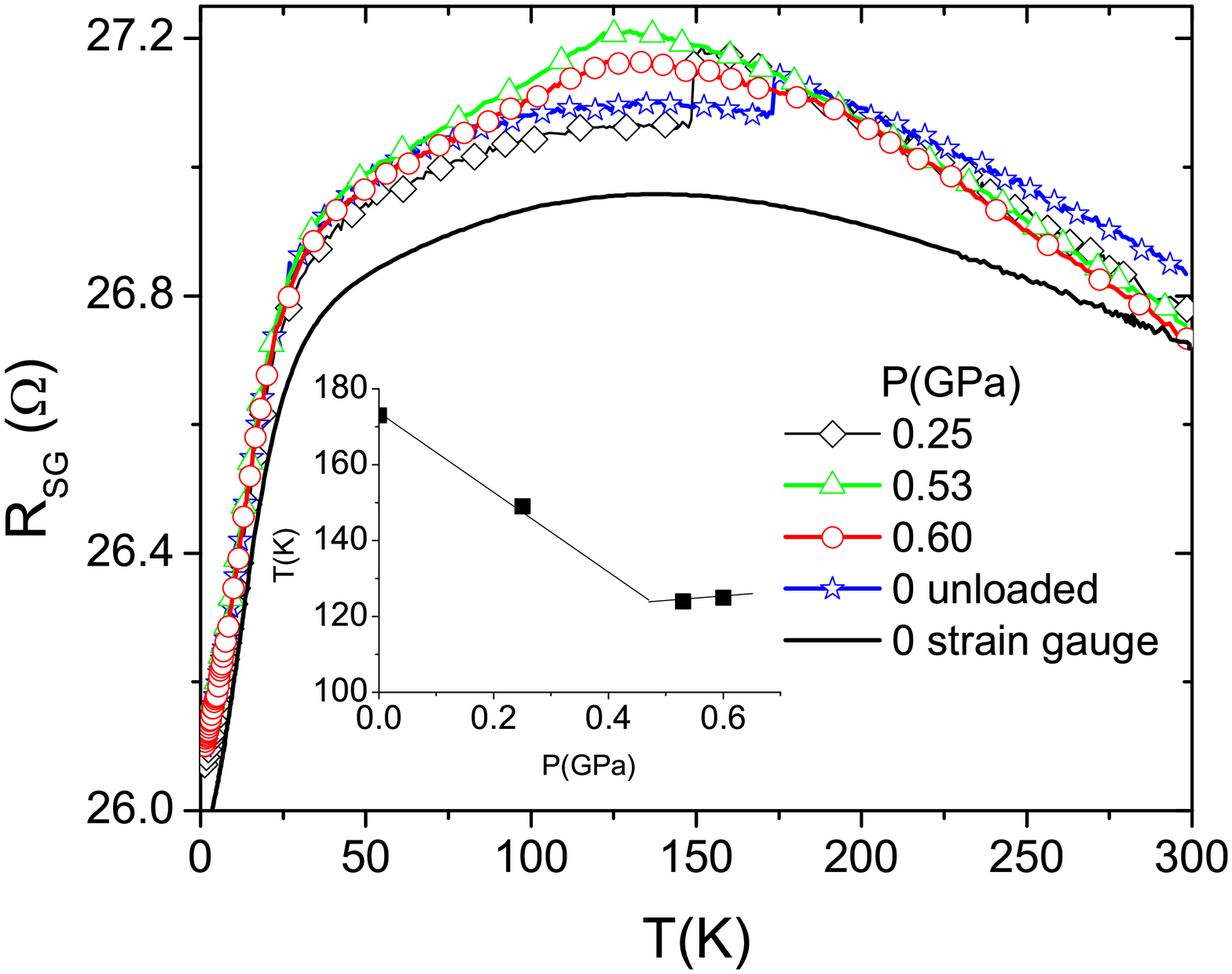}
\caption{(color online) Temperature dependent resistance of a constantan strain gauge glued along the tetragonal a-axis of CaFe$_2$As$_2$. One of nine data points is shown for clarity. The change in resistance is proportional to strain, as discussed in the text. For comparison, the solid line is plot of the temperature-dependent resistance of the free-standing strain gauge at P=0. Measurements are plotted for various increasing pressures, except for the P=0 data that were obtained on decreasing pressure from 0.6~GPa. Note the sharp jump in resistance at the structural transition for P=0 and 0.25~GPa but only a broad peak for pressures corresponding to range II in Fig.~2b. The inset is a plot of the structural transition temperature determined by these measurements, which are in good agreement with transition temperatures from resistivity measurements.}
\label{figure5}
\end{figure}

We interpret our observations as follows: At the phase~I/phase~II boundary, domains of cT phase begin to nucleate in the matrix of magnetic O domains, and the relative fraction of these domains reverses as the phase~II/phase~III boundary is approached. Though scattering at domain walls could account, in part, for the higher $\rho_0$ in phase~II relative to phase~III, the density of domain walls would need to be non-monotonic as a function of pressure to account for a maximum in $\rho_0$ shown in Fig.~4a. This seems unlikely. Alternatively, increased scattering in phase~II by dynamical processes provides a more plausible interpretation. The qualitatively lower resistivity over a broad temperature range in the pure cT phase implies a higher density of charge carriers than in the O structure, contrary to naive expectations from band structure calculations that find a reduced density of states in the cT structure~\cite{yildirim08a}. These 'doped' carriers from the cT phase are scattered by  fluctuations associated with the O structure. The origin of these fluctuations could be two-fold. In a strong coupling, local-moment picture of the magnetism, magnetic fluctuations are expected as the spin structure of O domains becomes increasingly frustrated due to a pressure-induced increase in As-As and Fe-As hybridization at pressures above 0.3~GPa.\cite{yildirim08, han08} On the other hand, in a weak coupling, spin-density-wave model, these changes in hybridization also will modify details of Fermi-surface topology and nesting conditions for the SDW, leading to spin fluctuations from nearly nested parts of the Fermi surface~\cite{stanev}. The correlation between an increase in $\rho_0$ and $T_c$ as well as the maximum in $-\Delta \chi_{ac}$ in phase~II is a consequence of the pressure dependence of the volume fraction of O and cT, where the O phase is the necessary source of magnetic fluctuations and the cT phase the source of carriers. The coupled spin-'doped' charge dynamics and bulk superconductivity, present only because of structural inhomogeneity, are absent in pure O and cT structures. Whether or not the domain structure is static or dynamic in phase~II cannot be established from these experiments, but expected pressure gradients at O/cT domain walls, due to significant structural differences of the phases ~\cite{kreyssig08}, could favor temporal fluctuations of the domains and Fe moments of the O structure ~\cite{mazin}. Without a small  pressure inhomogeneity inevitable with a soft but frozen pressure medium, it is not possible to nucleate the structural inhomogeneity that induces internal strain, which in turn stabilizes the composite structural phases.

The cT phase depends strongly on pressure history, as shown in Fig.~4b. Decreasing pressure from highest pressures locks in the low residual resistivity of phase~III to a pressure much below 0.8~GPa; there is no maximum in $\rho_0$ for $0.3 \leq P \leq 0.8$~GPa, and the maximum in $-\Delta \chi_{ac}$ appears near 0.3~GPa, all in contrast to results on increasing pressure. Recovery of the initial residual resistivity at ambient pressure suggests that the orthorhombic structure is the stable low temperature structure upon decreasing pressure, as concluded as well in recent structural studies~\cite{goldman082}. These conclusions also are consistent with the pressure dependence of $T_S$ determined on decreasing pressure (Fig.~2b). The strong pressure hysteresis of the boundary between phases~II and III and large thermal hysteresis at $T_S^h$ will induce structural and electronic inhomogeneities on decreasing pressure and further complicate an interpretation of $T_c(P)$ with decreasing pressure.

In summary, pressure-dependent resistivity and ac susceptibility measurements reveal strong coupling among superconductivity, structure and magnetism in CaFe$_2$As$_2$. The coexisting phase, which exists in a narrow window of increasing pressure, supports bulk superconductivity due to the coupled spin-charge dynamics special to the coexistence.

Work at Los Alamos was performed under the auspices of the U.S. Department of Energy/Office of Science and supported by the Los Alamos LDRD program. TP acknowledges support from KOSEF (2009-0058687) funded by the Korea government (MEST). VAS acknowledges support from the Russian Foundation for Basic Research (grant 09-02-00336).

\end{document}